\documentclass[prl,twocolumn,aps,epsf]{revtex4}
\usepackage{amssymb}
\usepackage{graphicx}
\usepackage{amsmath}
\usepackage{epstopdf}

\begin{document}

\title{Probing the helical edge states of a topological insulator\\
by Cooper-pair injection }
\author{P. Adroguer$^{(1)}$}
\author{C. Grenier$^{(1)}$}
\author{D. Carpentier$^{(1)}$}
\author{J. Cayssol$^{(2)}$}
\author{P. Degiovanni$^{(1)}$}
\author{E. Orignac$^{(1)}$}
\affiliation{$^{(1)}$Laboratoire de Physique, CNRS UMR5672, Ecole Normale Superieure de
Lyon, 46 All\'ee d'Italie, F-69364 Lyon Cedex 07, France}
\affiliation{$^{(2)}$CPMOH, CNRS UMR 5798, Universit\'e Bordeaux-I, 351, Cours de la
Lib\'eration, F-33045 Talence, France}

\begin{abstract}
We consider the proximity effect between a singlet s-wave superconductor and
the edge of a Quantum Spin Hall (QSH) topological insulator. We establish
that Andreev reflection at a QSH edge state/superconductor interface is
perfect while nonlocal Andreev processes through the superconductor are
totally suppressed. We compute the corresponding conductance and noise.
\end{abstract}

\maketitle

The prediction \cite{Bernevig2006} and the observation \cite{Konig2007,Roth2009} of the Quantum Spin Hall (QSH) state 
in mercury telluride (HgTe/CdTe) heterostructures have triggered a great deal of 
excitation in the condensed matter community \cite{Zhang2010,Moore2010,Hasan2010} since the QSH state realizes a two dimensional (2D) topologically ordered phase in 
the absence of magnetic field. The QSH state is distinguished from ordinary 
band insulators by the presence of a one-dimensional metal along its edge \cite{Kane2005}. Owing to the dominant role of the spin-orbit interaction, this edge state provides a unique strictly one-dimensional metal
where the spin is tied to the direction of motion of the carriers \cite{Wu2006}. This so-called 
helical property and the associated time-reversal symmetry imply the absence of backscattering on 
non magnetic impurities. 

So far the existence of the helical liquid has been confirmed by multiterminal transport
measurements performed with normal leads \cite{Konig2007,Roth2009}. Since the QSH state 
exists under zero magnetic field, in contrast to
the integer and fractional quantum Hall states, it can also be
probed by the powerful methods of superconducting proximity effect \cite{Tinkham1996}. Along these 
lines, Andreev spectroscopy has been recently suggested to
characterize the quasi-relativistic dynamics of 2D bulk carriers in
doped HgTe/CdTe quantum wells \cite{Guigou2010}. Furthermore helical liquids might also 
be useful to analyze the entanglement of electrons injected from a singlet $s-$wave superconductor \cite{Choi2010,Loss2010}.

\begin{figure}[tbp]
\centerline{
    \includegraphics[width=8cm]{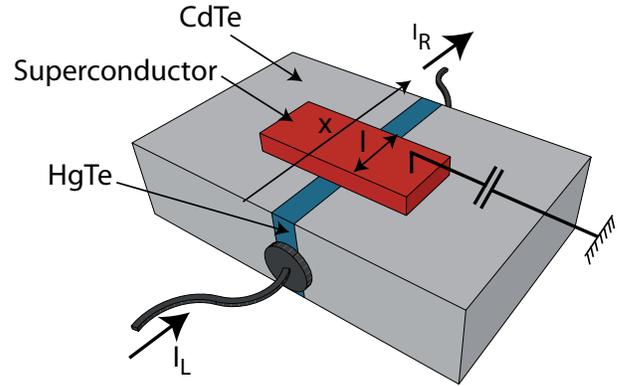} }  
\caption{(Color online). Schematic representation of the proposed experimental setup. The
Quantum Spin Hall phase is realized in an inverted and insulating HgTe/CdTe
quantum well. Transport along the one-dimensional edge of the QSH phase is
measured by a standard two terminals setup with normal electrodes. Between
these two electrodes, a superconducting electrode is deposited over a length 
$l$ on one side of the sample. }
\label{fig:setup3D}
\end{figure}

In this Rapid Communication, we theoretically investigate the edge transport of a Quantum Spin Hall insulator 
in presence of a single superconducting probe. As a result of helicity conservation and 
absence of backscattering channel, we find that an electron can be either Andreev reflected as a hole, or transmitted 
as an electron. In a standard metal or in a carbon nanotube, there would be two additional possibilities whereby 
the electron can be reflected as an electron, or transmitted as a hole \cite{Beckmann2004,Cadden2006,Hofstetter2009,Herrmann2010,Byers1995,Hartog1996,Deutscher2000}. We compute the conductance and the noise 
associated to this partitioning in two outgoing channels, 
instead of four channels in standard 1D metals. The related experiments could be implemented readily using a side superconductor 
contacted to current HgTe/CdTe samples \cite{Konig2007,Roth2009}. Our results also apply to other possible experimental 
realizations including the recently proposed inverted type II semiconductor quantum wells \cite{LiuHughes2008} and ultrathin Bi$_2$Se$_3$ films \cite{Liu2010,Lu2010}. Finally we contrast our results with Andreev
transport through neutral Majorana fermions as realized at a triple
interface between a ferromagnet, a superconductor and a topological
insulator \cite{Fu2008,Nilsson2008,Akhmerov2009,Fu2009,Law2009b,TanakaYokoyama2009}.

In our proposed setup, a superconducting probe is deposited near an inverted
HgTe/CdTe quantum well thereby inducing superconducting correlations within
the QSH edge state (Fig.~\ref{fig:setup3D}). The counterpropagating
electrons or holes are detected by distant normal metallic contacts. We
assume a wide enough HgTe well so that scattering between opposite edges is absent \cite%
{Zhou2008}. The opposite limit of strong inter-edge scattering has been
addressed in Refs.\cite{Choi2010,Loss2010}. In the absence of
superconductivity, the single pair of gapless edge states is described by
the one-dimensional Dirac Hamiltonian%
\begin{equation}
H_{0}=-i\hbar v_{F}\int_{-\infty }^{\infty }dx~\left( \psi _{\uparrow
}^{\dagger }\partial _{x}\psi _{\uparrow }-\psi _{\downarrow }^{\dagger
}\partial _{x}\psi _{\downarrow }\right) ,  \label{eq:qshe-hamiltonian}
\end{equation}%
where $h=2\pi \hbar $ is Planck's constant and $v_{F}$ the Fermi
velocity. Without any loss of generality, we have chosen the convention that
the (pseudo)spin-up electrons associated with field operator $\psi
_{\uparrow }(x)$ are right moving while the spin-down electrons are
left moving. In contrast to a usual metal, there are no right movers with
down spin or left movers with up spins. As a result, in a QSH edge, the
product of the spin by the velocity is always positive which is called
helicity conservation. These left and right movers are only well defined inside 
the bulk gap of the insulator which is typically $E_{g} \sim 1-30$ meV in HgTe/CdTe quantum wells \cite{Konig2007,Roth2009}. 

We further assume that the superconductor induces a
gap $\Delta(x)$ over a finite length $l$ of the helical liquid \cite%
{Fu2008,Nilsson2008,Akhmerov2009,Fu2009,Law2009b,TanakaYokoyama2009}. The edge transport is
then described by the effective Hamiltonian 
\begin{equation}  \label{eq:proximity}
H=H_{0}+\int_{0}^{l} dx~ \left(\Delta^*(x)\psi _{\downarrow
}(x)\psi_{\uparrow }(x)+\text{H.c.} \right),
\end{equation}%
where the amplitude of the proximity induced 
gap depends upon the coupling between the edge and the superconductor \cite{Sarma2010}. The induced gap amplitude $\left\vert \Delta \right\vert$ may reach 
at best the intrinsic gap of the superconductor, namely $\left\vert \Delta \right\vert \sim 0.1-1$ meV when using aluminium or niobium.

\medskip

We first discuss qualitatively the available scattering processes in the
opposite limits of long ($l \gg \xi=\hbar v_{F}/\left\vert \Delta \right\vert
$) and short ($l\lesssim \xi \sim 10-100$ nm) superconductor using only helicity
conservation and time-reversal symmetry.

An electron with energy $\varepsilon <|\Delta| $ cannot be transmitted
through a long superconductor ($l\gg \xi$) since the penetration depth of
the evanescent Bogoliubov quasiparticle in the superconductor is set by its coherence
length $\xi$. Hence in a standard metal, this incident electron can be either
reflected as an electron (electron backscattering) or as a hole (Andreev
reflection) at a single normal-superconducting (NS) metal interface \cite{Andreev1964}. Electron reflection
changes the direction of propagation while conserving the spin: it is thus
forbidden by helicity conservation in a QSH edge state. Due to unitarity,
this absence of electronic backscattering implies Andreev reflection with
unit probability even in presence of disorder and/or potential barrier at
the interface. Such a perfect Andreev reflection is very difficult to
achieve in standard metals where any defect or material parameters mismatch
will induce a sizeable electron backscattering \cite{Blonder1982}.

Interestingly, another kind of perfect Andreev reflection has been predicted
recently for a Fermi lead coupled to a Majorana fermion at its end \cite%
{Law2009b}. This perfect Andreev reflection results from the self conjugate
property of the Majorana fermion which couples the electron and hole modes
with equal amplitude. Nevertheless this resonant Andreev reflection requires
a matching between the energy of the incident electron and the energy of the
Majorana mode. By contrast, in our setup, the perfect Andreev reflection is
achieved for all energies below the superconducting gap and relies only on
the helical property of the lead.

Moreover, state-of-the art nanolithography allows for the
realization of narrower superconducting regions ($l\lesssim \xi$) covering a
metal strip \cite{Beckmann2004,Cadden2006} or even a single carbon nanotube
or nanowire \cite{Hofstetter2009,Herrmann2010}. In such a
normal-superconducting-normal (NSN) geometry, subgap quasiparticles can also
be transmitted. In a Fermi liquid lead, an incoming electron can be either
transmitted as an electron (elastic cotunneling) or as a hole (non local
Andreev process) \cite%
{Byers1995,Hartog1996,Deutscher2000}. The non local
Andreev process have been evidenced recently in several experiments \cite%
{Hofstetter2009,Herrmann2010}. Within the QSH edge, such Andreev
transmission is again strictly forbidden by helicity conservation.
Interestingly, in the presence of Majorana fermions, the Andreev transmission is
restored and dominates the normal tunneling \cite{Law2009b,Nilsson2008}.

\begin{figure}[tbp]
\centerline{
    \includegraphics[width=9cm]{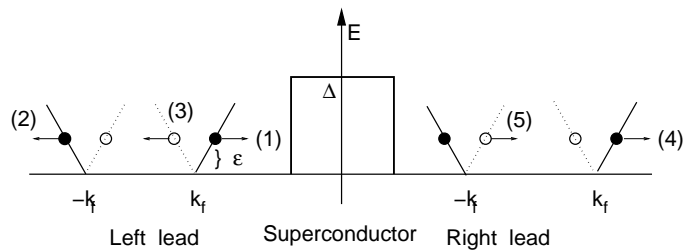} }
\caption{(Color online). Scattering processes. With Fermi liquid leads, an incident electron
(1) can be backscattered as an electron (2), reflected as a hole (local
Andreev reflection) (3), transmitted as an electron (4) or transmitted as a
hole (non local Andreev process) (5). In the QSH edge state, helicity
conservation prevents electronic backscattering (2) and hole transmission
(5). Furthermore, at low energy and for a wide superconductor, electron
transmission (4) vanishes and only Andreev reflection (3) remains.}
\label{fig:scattering}
\end{figure}

Therefore along a QSH edge state, an incident electron can be only reflected as a hole by a
superconducting barrier, or transmitted as an electron through it (Fig.~\ref%
{fig:scattering}). Using the
Landauer-B\"{u}ttiker formalism, we now provide the quantitative theory of
this partitioning between Andreev reflection and normal transmission which
holds for arbitrary superconductor length $l$, and energy $\epsilon$. We
define the incoming fields, $\psi _{\uparrow ,i}= \psi _{\uparrow }(x=0, t)$
and $\psi^\dagger _{\downarrow ,i} = \psi^\dagger _{\downarrow}(x=l,t)$, in
terms of the fermionic operators $\psi
_{\uparrow,\downarrow}(x,t)=\psi_{\uparrow,\downarrow}(0,t \mp x/v_{F})$
which capture the ballistic helical propagation within the edge state. Since
only normal transmission and Andreev reflection are allowed the outgoing
fields are defined as $\psi_{\uparrow ,o}=\psi _{\uparrow }(x=l, t)$ and $%
\psi^\dagger _{\downarrow ,o}= \psi^\dagger _{\downarrow}(x=0,t)$. The
quasiparticle energy $\epsilon$ being conserved, it is convenient to
introduce the following Fourier representation: 
\begin{eqnarray}  \label{eq:fourier}
\psi_\sigma(x,t)=\int_{-\infty }^{\infty } \frac{d\epsilon}{2\pi\hbar} e^{-i\epsilon t/\hbar}
\psi_\sigma(x,\epsilon).
\end{eqnarray}

Considering the solutions of the Dirac equation outside the barrier, and
applying time reversal symmetry, we obtain: 
\begin{subequations}
\label{eq:solutions}
\begin{align}  \label{eq:sol-down}
\psi_{\uparrow,o} (\epsilon) &= t(\epsilon) \psi_{\uparrow,i}(\epsilon) -%
\frac{r^*(\epsilon)t(\epsilon)}{t^*(\epsilon)} \psi^\dagger_{\downarrow,i}(%
-\epsilon)\, , \\
\psi^\dagger_{\downarrow,o} (-\epsilon) &= r(\epsilon)
\psi_{\uparrow,i}(\epsilon) + t(\epsilon) \psi^\dagger_{\downarrow,i}
(-\epsilon)\,.
\end{align}
 
The scattering coefficients $r(\epsilon),t(\epsilon)$  which relate the incoming chiral fermionic fields to the
outgoing ones must be obtained from the full solution of the one-dimensional
Dirac equation associated with Eqs.(\ref{eq:qshe-hamiltonian},\ref{eq:proximity}). The probability for an electron of energy $\epsilon$ (with
respect to the superconductor chemical potential) to be transmitted through
the superconducting barrier is $T_\epsilon=|t(\epsilon)|^2$ and  the
probability for an electron of spin $\sigma$ and energy $\epsilon$ to be
reflected as a hole of the same energy on the $-\sigma$ spin branch is $%
R_\epsilon=|r(\epsilon)|^2$. Current conservation always imposes $%
R_\epsilon+T_\epsilon=1$ irrespective of the specific shape of the pairing potential $\Delta(x)$. %

\begin{figure}[tbp]
\includegraphics[width=8cm]{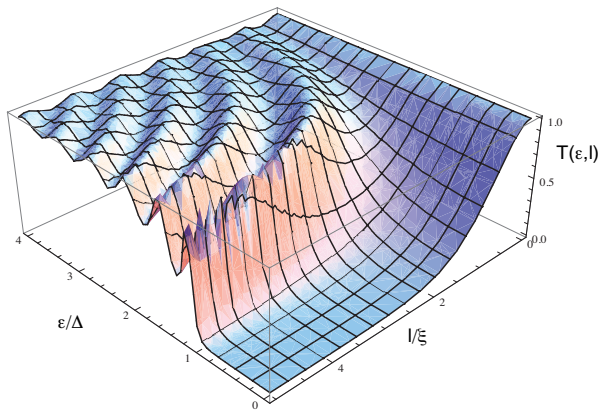}
\caption{(Color online). Plot of $T_\epsilon$ for a rectangular gap function $\Delta(x)=\Delta 
\protect\theta(x) \protect\theta(l-x)$: the horizontal axis display $l/\protect\xi$ and
$\protect\epsilon /|\Delta|$. For $l\gtrsim \protect\xi$,
the transmission vanishes for subgap electrons $\protect\epsilon%
\leq |\Delta|$. Oscillations of the transmission coefficient for $\protect%
\epsilon \geq |\Delta|$ can be interpreted as Fabry-Perot resonances of
the Bogoliubov excitations through the superconducting barrier.}
\label{fig:transmission}
\end{figure}

Moreover in long superconducting segments, $l\gtrsim \xi$, electrons satisfying $%
\epsilon\leq |\Delta|$ experience total local Andreev reflection ($%
R_\epsilon=1$) at each interface whereas for shorter superconducting
regions, a finite electronic transmission is possible. On the other hand and
for any length $l$, electrons of very high energy $\epsilon \gg |\Delta|$ are perfectly transmitted (%
$T_\epsilon=1$) as pure electron-like quasiparticles through the
superconducting barrier. For intermediate energies $\epsilon \gtrsim |\Delta|$, Bogoliubov
quasiparticles experience Fabry-Perot like transmission resonances at
discrete energies $\bar{\epsilon}_n$ given by condition $r(\bar{\epsilon}%
_n)=0$. In the case of a rectangular barrier $\Delta(x)=\Delta \theta(x)
\theta(l-x)$, those resonances are located at $\hbar \bar{\epsilon}_n=\sqrt{%
\Delta^2 + (\hbar v_F n \pi/l)^2}$ with $n$ integer (Fig.~\ref%
{fig:transmission}).

\medskip

We now compute the conductance and noise of the three terminal
setup when the left and right normal leads are biased at the respective
potentials $V_{L},V_{R}$ while the superconductor is grounded (Fig.~\ref%
{fig:setup3D}).  The condition that the electrons incoming from the
reservoir are in thermal equilibrium is expressed as: 
 $\langle \psi^\dagger_{\sigma,i}(\epsilon)
\psi_{\sigma,i}(\epsilon')\rangle=\frac{2\pi\hbar}{ v_F}
\delta(\epsilon-\epsilon') n_\sigma(\epsilon)$ 
and $\langle \psi_{\sigma,i}(\epsilon)
\psi_{\sigma,i}^\dagger(\epsilon')\rangle=\frac{2\pi \hbar}{ v_F}
\delta(\epsilon-\epsilon') (1-n_\sigma(\epsilon))$ 
where $n_{\uparrow /\downarrow }(\epsilon )$ is the Fermi-Dirac
distribution function for the $\uparrow $ (resp. $\downarrow $) incoming
electrons, with chemical potential  $\mu _{L}=eV_{L}$ (resp. $\mu
_{R}=eV_{R}$). Along the edge state, the current operator is defined as $I(x,t)=-ev_{F}\left( \psi
_{\uparrow }^{\dagger }\psi _{\uparrow }-\psi _{\downarrow }^{\dagger }\psi
_{\downarrow }\right) $. The current injected from the superconductor is described 
by the operator $I_S(t)=I_R(t)-I_L(t)$, where $I_L(t)=I(x=0,t)$ and $I_R(t)=I(x=l,t)$ are the currents 
flowing in the left and right normal leads respectively (Fig.~\ref{fig:setup3D}). Using 
Eqs.~(\ref{eq:solutions}), the average current injected
by the superconductor is found to be: 
\end{subequations}
\begin{equation}
\langle I_{S}\rangle=\frac{2e}{h}\int_{-\infty }^{\infty } R_\epsilon \,(n_{\uparrow }(\epsilon
)+n_{\downarrow }(-\epsilon )-1)\,d\epsilon \,\,.  \label{eq:current-super}
\end{equation}%
 When $\mu _{R}=\mu
_{L}=eV$, Eq.~(\ref{eq:current-super}) leads to a differential conductance
given by: $(\partial \langle I_{S} \rangle/\partial V)_{V=0}=(4e^{2}/h)\times R_{\epsilon
=0}$ increasing with $l$ from zero (no coupling to the superconductor for $%
l=0$) to $4e^{2}/h$ in the $l\gg \xi $ limit. Therefore we predict that the conductance 
$(\partial \langle I_{S} \rangle/\partial V)({V,T})$ must saturate at $4e^{2}/h$ for low voltage/temperature (max$(eV,k_{B}T) \ll \Delta$, $k_{B}$ being the Boltzmann
constant) which is the expected value for two perfectly Andreev reflecting N/S interfaces in parallel. 

The noise power $S_{S} (\omega)$ of the current injected from the superconductor, i.e. the Fourier transform of the autocorrelator $\langle
{I_{S}(0)I_{S}(t)}\rangle $, can be
computed from the scattering formalism with the help of Wick's
theorem. At zero frequency, we find:  
\begin{eqnarray}  \label{eq:noise}
S_{S}(0) & = & \frac{8e^2}{h}\int_{-\infty }^{\infty }
R_\epsilon\left(n_{\uparrow}(1-n_{\uparrow})+n_{\downarrow}(1-n_{%
\downarrow})\right)\,d\epsilon  \notag \\
&+&\frac{8e^2}{h}\int_{-\infty }^{\infty } R_\epsilon\,T_\epsilon
\left(n_{\uparrow}+n_{\downarrow}-1 \right)^2d\epsilon \, ,
\end{eqnarray}
where the shorthand notations $n_{\uparrow}=n_{\uparrow}(\epsilon)$ and $n_{\downarrow}=n_{\downarrow}(-\epsilon)$ are used. The first term in Eq.~(\ref{eq:noise}) originates from the equilibrium thermal noise of the reservoirs whereas the second term is the nonequilibrium contribution to the superconducting current noise coming from the Andreev reflection/normal transmission partitioning. In the vanishing voltage limit ($V \rightarrow 0$) only the first line of Eq.~(\ref{eq:noise}) contributes to the noise, and the Johnson-Nyquist 
relation $S_{S}(0)=4 k_{B} T (\partial \langle I_{S} \rangle/\partial V)_{V=0}$ is satisfied. 

\begin{figure}[tbp]
\includegraphics[width=8cm]{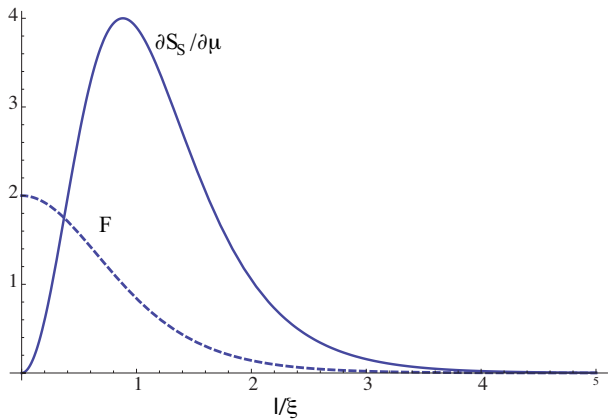}
\caption{(Color online). The noise response $\partial S_{S}/\partial \protect\mu $ in units of $%
e^{2}/h$ (solid line) is maximal around $l \sim \xi$. The Fano factor $F=S_{S}/2e\langle I_{S}\rangle $ (dashed line) as
a function of $l/\protect\xi $.}
\label{fig:noise-response}
\end{figure}

At zero temperature, only partially transmitted electrons ($T_\epsilon\neq 0,1$) will
generate a finite noise. For a rectangular pair potential $\Delta(x)=\Delta \theta(x)
\theta(l-x)$, the noise response to a low bias $eV _{L}=eV _{R}=\mu$, 
\begin{equation}
\frac{\partial S_{S}}{\partial \mu }=\frac{16e^{2}}{h}\,\frac{\sinh ^{2}{%
(l/\xi )}}{\cosh ^{4}{(l/\xi )}},  \label{eq:noise-deriv}
\end{equation}%
reaches the maximal value of $4e^{2}/h$ for a superconductor width corresponding to an equal 
partitioning between the local Andreev reflection and the normal transmission processes, namely $%
T_{\epsilon =0}=R_{\epsilon =0}=1/2$ (Fig.~\ref{fig:noise-response}, solid line). At
zero temperature and vanishing bias voltage $V\rightarrow 0$, the
corresponding Fano factor $F=S_{S}/2e\langle I_{S}\rangle =2T_{\epsilon =0}$ is twice the transmission probability 
through the barrier (Fig.~\ref{fig:noise-response}, dashed line), in agreement with 
the transfer of charges $2e$ between the superconductor and the QSH edge. For long superconducting 
segments, $l\gtrsim \xi$, one obtains two uncorrelated and noiseless QSH/superconductor interfaces where for 
subgap electrons the quasiparticle currents 
are converted into supercurrent through perfect normal Andreev reflection.

\medskip

In conclusion, we have considered a junction between a superconductor
electrode and a QSH phase. The conservation of helicity and the resulting
absence of backscattering manifest themselves into a perfect Andreev reflection
for long junction in the subgap regime. This leads to unique signatures of the helical nature of
the edge states, including in particular the existence of noiseless injected
currents on both sides of long QSH/S/QSH junctions. Our model assumed that the QSH edge states were not 
reconstructed in the presence of the superconductor. In particular, we have neglected the possibility of the formation of new QSH edge channels or 
a local 2D metallic puddle beneath the superconductor. Such non-universal effects might prevent the observation of full Andreev reflection. However, their 
consideration in a selfconsistent Bogoliubov-de Gennes theory, which depends on the full bandstructures of both the superconductor and the 
topological insulator, is beyond the scope of this Rapid Communication.

Finally we further anticipate interesting two-particle interference effects in the high frequency regime. In particular a 
fermionic version of the Hong-Ou-Mandel experiment \cite{Ou1987} could be implemented using
single electron and single hole excitations obtained from Lorentzian voltage
pulses applied to the left and right reservoirs \cite{Levitov1996}.

\medskip

\begin{acknowledgments}
J.C. thanks Bjoern Trauzettel, Patrik Recher and Marine Guigou for stimulating
discussions. J.C. acknowledges funding from the Institut de
Physique Fondamentale in Bordeaux. This work was supported by the Agence Nationale de la Recherche
under Grant No. ANR-07-NANO-011: ELEC-EPR (J.C.) and under the program blanc ISOTOP (D.C., J.C. and
E.O.).
\end{acknowledgments}

\bibliographystyle{apsrev}
\bibliography{qshe}

\begin{thebibliography}{34}
\expandafter\ifx\csname natexlab\endcsname\relax\def\natexlab#1{#1}\fi
\expandafter\ifx\csname bibnamefont\endcsname\relax
  \def\bibnamefont#1{#1}\fi
\expandafter\ifx\csname bibfnamefont\endcsname\relax
  \def\bibfnamefont#1{#1}\fi
\expandafter\ifx\csname citenamefont\endcsname\relax
  \def\citenamefont#1{#1}\fi
\expandafter\ifx\csname url\endcsname\relax
  \def\url#1{\texttt{#1}}\fi
\expandafter\ifx\csname urlprefix\endcsname\relax\def\urlprefix{URL }\fi
\providecommand{\bibinfo}[2]{#2}
\providecommand{\eprint}[2][]{\url{#2}}

\bibitem[{\citenamefont{Bernevig et~al.}(2006)\citenamefont{Bernevig, Hughes,
  and Zhang}}]{Bernevig2006}
\bibinfo{author}{\bibfnamefont{B.}~\bibnamefont{Bernevig}},
  \bibinfo{author}{\bibfnamefont{T.~L.} \bibnamefont{Hughes}},
  \bibnamefont{and} \bibinfo{author}{\bibfnamefont{S.-C.} \bibnamefont{Zhang}},
  \bibinfo{journal}{Science} \textbf{\bibinfo{volume}{314}},
  \bibinfo{pages}{1757} (\bibinfo{year}{2006}).

\bibitem[{\citenamefont{Konig et~al.}(2007)\citenamefont{Konig, Wiedmann,
  Brune, Roth, Buhmann, Molenkamp, Qi, and Zhang}}]{Konig2007}
\bibinfo{author}{\bibfnamefont{M.}~\bibnamefont{Konig}},
  \bibinfo{author}{\bibfnamefont{S.}~\bibnamefont{Wiedmann}},
  \bibinfo{author}{\bibfnamefont{C.}~\bibnamefont{Brune}},
  \bibinfo{author}{\bibfnamefont{A.}~\bibnamefont{Roth}},
  \bibinfo{author}{\bibfnamefont{H.}~\bibnamefont{Buhmann}},
  \bibinfo{author}{\bibfnamefont{L.~W.} \bibnamefont{Molenkamp}},
  \bibinfo{author}{\bibfnamefont{X.-L.} \bibnamefont{Qi}}, \bibnamefont{and}
  \bibinfo{author}{\bibfnamefont{S.-C.} \bibnamefont{Zhang}},
  \bibinfo{journal}{Science} \textbf{\bibinfo{volume}{318}},
  \bibinfo{pages}{766} (\bibinfo{year}{2007}).

\bibitem[{\citenamefont{Roth et~al.}(2009)\citenamefont{Roth, Bruene, Buhmann,
  Molenkamp, Maciejko, Qi, and Zhang}}]{Roth2009}
\bibinfo{author}{\bibfnamefont{A.}~\bibnamefont{Roth}},
  \bibinfo{author}{\bibfnamefont{C.}~\bibnamefont{Bruene}},
  \bibinfo{author}{\bibfnamefont{H.}~\bibnamefont{Buhmann}},
  \bibinfo{author}{\bibfnamefont{L.}~\bibnamefont{Molenkamp}},
  \bibinfo{author}{\bibfnamefont{J.}~\bibnamefont{Maciejko}},
  \bibinfo{author}{\bibfnamefont{X.-L.} \bibnamefont{Qi}}, \bibnamefont{and}
  \bibinfo{author}{\bibfnamefont{S.-C.} \bibnamefont{Zhang}},
  \bibinfo{journal}{Science} \textbf{\bibinfo{volume}{325}},
  \bibinfo{pages}{294} (\bibinfo{year}{2009}).

\bibitem[{\citenamefont{Qi and Zhang}(2010)}]{Zhang2010}
\bibinfo{author}{\bibfnamefont{X.-L.} \bibnamefont{Qi}} \bibnamefont{and}
  \bibinfo{author}{\bibfnamefont{S.-C.} \bibnamefont{Zhang}},
  \bibinfo{journal}{Physics Today} \textbf{\bibinfo{volume}{63}},
  \bibinfo{pages}{33} (\bibinfo{year}{2010}).

\bibitem[{\citenamefont{Moore}(2010)}]{Moore2010}
\bibinfo{author}{\bibfnamefont{J.~E.} \bibnamefont{Moore}},
  \bibinfo{journal}{Nature} \textbf{\bibinfo{volume}{464}},
  \bibinfo{pages}{194} (\bibinfo{year}{2010}).

\bibitem[{\citenamefont{Hasan and Kane}(2010)}]{Hasan2010}
\bibinfo{author}{\bibfnamefont{M.~Z.} \bibnamefont{Hasan}} \bibnamefont{and}
  \bibinfo{author}{\bibfnamefont{C.~L.} \bibnamefont{Kane}}
  (\bibinfo{year}{2010}), \bibinfo{note}{arXiv:1002.3895}.

\bibitem[{\citenamefont{Kane and Mele}(2005)}]{Kane2005}
\bibinfo{author}{\bibfnamefont{C.~L.} \bibnamefont{Kane}} \bibnamefont{and}
  \bibinfo{author}{\bibfnamefont{E.~J.} \bibnamefont{Mele}},
  \bibinfo{journal}{Phys. Rev. Lett.} \textbf{\bibinfo{volume}{95}},
  \bibinfo{pages}{226801} (\bibinfo{year}{2005}).

\bibitem[{\citenamefont{Wu et~al.}(2006)\citenamefont{Wu, Bernevig, and
  Zhang}}]{Wu2006}
\bibinfo{author}{\bibfnamefont{C.}~\bibnamefont{Wu}},
  \bibinfo{author}{\bibfnamefont{B.}~\bibnamefont{Bernevig}}, \bibnamefont{and}
  \bibinfo{author}{\bibfnamefont{S.-C.} \bibnamefont{Zhang}},
  \bibinfo{journal}{Phys. Rev. Lett.} \textbf{\bibinfo{volume}{96}},
  \bibinfo{pages}{106401} (\bibinfo{year}{2006}).

\bibitem[{\citenamefont{Tinkham}(1996)}]{Tinkham1996}
\bibinfo{author}{\bibfnamefont{M.}~\bibnamefont{Tinkham}},
  \emph{\bibinfo{title}{Introduction to Superconductivity}}
  (\bibinfo{publisher}{McGraw-Hill New York}, \bibinfo{year}{1996}).

\bibitem[{\citenamefont{Guigou and Cayssol}()}]{Guigou2010}
\bibinfo{author}{\bibfnamefont{M.}~\bibnamefont{Guigou}} \bibnamefont{and}
  \bibinfo{author}{\bibfnamefont{J.}~\bibnamefont{Cayssol}},
  \bibinfo{note}{arXiv:1005.1055}.

\bibitem[{\citenamefont{Choi}()}]{Choi2010}
\bibinfo{author}{\bibfnamefont{M.-S.} \bibnamefont{Choi}},
  \bibinfo{note}{arXiv:1002.3144}.

\bibitem[{\citenamefont{Sato et~al.}()\citenamefont{Sato, Loss, and
  Tserkovnyak}}]{Loss2010}
\bibinfo{author}{\bibfnamefont{K.}~\bibnamefont{Sato}},
  \bibinfo{author}{\bibfnamefont{D.}~\bibnamefont{Loss}}, \bibnamefont{and}
  \bibinfo{author}{\bibfnamefont{Y.}~\bibnamefont{Tserkovnyak}},
  \bibinfo{note}{arXiv:1003.4316}.

\bibitem[{\citenamefont{Beckmann et~al.}(2004)\citenamefont{Beckmann, Weber,
  and {L\"ohneysen}}}]{Beckmann2004}
\bibinfo{author}{\bibfnamefont{D.}~\bibnamefont{Beckmann}},
  \bibinfo{author}{\bibfnamefont{H.}~\bibnamefont{Weber}}, \bibnamefont{and}
  \bibinfo{author}{\bibfnamefont{H.}~\bibnamefont{{L\"ohneysen}}},
  \bibinfo{journal}{Phys. Rev. Lett.} \textbf{\bibinfo{volume}{93}},
  \bibinfo{pages}{197003} (\bibinfo{year}{2004}).

\bibitem[{\citenamefont{Cadden-Zimansky and Chandrasekhar}(2006)}]{Cadden2006}
\bibinfo{author}{\bibfnamefont{P.}~\bibnamefont{Cadden-Zimansky}}
  \bibnamefont{and}
  \bibinfo{author}{\bibfnamefont{V.}~\bibnamefont{Chandrasekhar}},
  \bibinfo{journal}{Phys. Rev. Lett.} \textbf{\bibinfo{volume}{97}},
  \bibinfo{pages}{237003} (\bibinfo{year}{2006}).

\bibitem[{\citenamefont{Hofstetter et~al.}(2009)\citenamefont{Hofstetter,
  Csonka, Nygard, and Schonenberger}}]{Hofstetter2009}
\bibinfo{author}{\bibfnamefont{L.}~\bibnamefont{Hofstetter}},
  \bibinfo{author}{\bibfnamefont{S.}~\bibnamefont{Csonka}},
  \bibinfo{author}{\bibfnamefont{J.}~\bibnamefont{Nygard}}, \bibnamefont{and}
  \bibinfo{author}{\bibfnamefont{C.}~\bibnamefont{Schonenberger}},
  \bibinfo{journal}{Nature (London)} \textbf{\bibinfo{volume}{461}},
  \bibinfo{pages}{960} (\bibinfo{year}{2009}).

\bibitem[{\citenamefont{Herrmann et~al.}(2010)\citenamefont{Herrmann, Portier,
  Roche, Yeyati, Kontos, and Strunk}}]{Herrmann2010}
\bibinfo{author}{\bibfnamefont{L.~G.} \bibnamefont{Herrmann}},
  \bibinfo{author}{\bibfnamefont{F.}~\bibnamefont{Portier}},
  \bibinfo{author}{\bibfnamefont{P.}~\bibnamefont{Roche}},
  \bibinfo{author}{\bibfnamefont{A.~L.} \bibnamefont{Yeyati}},
  \bibinfo{author}{\bibfnamefont{T.}~\bibnamefont{Kontos}}, \bibnamefont{and}
  \bibinfo{author}{\bibfnamefont{C.}~\bibnamefont{Strunk}},
  \bibinfo{journal}{Phys. Rev. Lett.} \textbf{\bibinfo{volume}{104}},
  \bibinfo{pages}{026801} (\bibinfo{year}{2010}).

\bibitem[{\citenamefont{Byers and {Flatt\'e}}(1995)}]{Byers1995}
\bibinfo{author}{\bibfnamefont{J.}~\bibnamefont{Byers}} \bibnamefont{and}
  \bibinfo{author}{\bibfnamefont{M.}~\bibnamefont{{Flatt\'e}}},
  \bibinfo{journal}{Phys. Rev. Lett.} \textbf{\bibinfo{volume}{74}},
  \bibinfo{pages}{306} (\bibinfo{year}{1995}).

\bibitem[{\citenamefont{{den Hartog} et~al.}(1996)\citenamefont{{den Hartog},
  Kapteyn, van Wees, and Klapwijk}}]{Hartog1996}
\bibinfo{author}{\bibfnamefont{S.~G.} \bibnamefont{{den Hartog}}},
  \bibinfo{author}{\bibfnamefont{C.~M.~A.} \bibnamefont{Kapteyn}},
  \bibinfo{author}{\bibfnamefont{B.~J.} \bibnamefont{van Wees}},
  \bibnamefont{and} \bibinfo{author}{\bibfnamefont{T.}~\bibnamefont{Klapwijk}},
  \bibinfo{journal}{Phys. Rev. Lett.} \textbf{\bibinfo{volume}{77}},
  \bibinfo{pages}{4954} (\bibinfo{year}{1996}).

\bibitem[{\citenamefont{Deutscher and Feinberg}(2000)}]{Deutscher2000}
\bibinfo{author}{\bibfnamefont{G.}~\bibnamefont{Deutscher}} \bibnamefont{and}
  \bibinfo{author}{\bibfnamefont{D.}~\bibnamefont{Feinberg}},
  \bibinfo{journal}{Appl. Phys. Lett.} \textbf{\bibinfo{volume}{76}},
  \bibinfo{pages}{487} (\bibinfo{year}{2000}).

\bibitem[{\citenamefont{Liu et~al.}(2008)\citenamefont{Liu, Hughes, Qi, Wang,
  and Zhang}}]{LiuHughes2008}
\bibinfo{author}{\bibfnamefont{C.}~\bibnamefont{Liu}},
  \bibinfo{author}{\bibfnamefont{T.}~\bibnamefont{Hughes}},
  \bibinfo{author}{\bibfnamefont{X.-L.} \bibnamefont{Qi}},
  \bibinfo{author}{\bibfnamefont{K.}~\bibnamefont{Wang}}, \bibnamefont{and}
  \bibinfo{author}{\bibfnamefont{S.-C.} \bibnamefont{Zhang}},
  \bibinfo{journal}{Phys. Rev. Lett.} \textbf{\bibinfo{volume}{100}},
  \bibinfo{pages}{236601} (\bibinfo{year}{2008}).

\bibitem[{\citenamefont{Liu et~al.}(2010)\citenamefont{Liu, Zhang, Yan, Qi,
  Frauenheim, Dai, Fang, and Zhang}}]{Liu2010}
\bibinfo{author}{\bibfnamefont{C.-X.} \bibnamefont{Liu}},
  \bibinfo{author}{\bibfnamefont{H.}~\bibnamefont{Zhang}},
  \bibinfo{author}{\bibfnamefont{B.}~\bibnamefont{Yan}},
  \bibinfo{author}{\bibfnamefont{X.-L.} \bibnamefont{Qi}},
  \bibinfo{author}{\bibfnamefont{T.}~\bibnamefont{Frauenheim}},
  \bibinfo{author}{\bibfnamefont{X.}~\bibnamefont{Dai}},
  \bibinfo{author}{\bibfnamefont{Z.}~\bibnamefont{Fang}}, \bibnamefont{and}
  \bibinfo{author}{\bibfnamefont{S.-C.} \bibnamefont{Zhang}},
  \bibinfo{journal}{Phys. Rev. B} \textbf{\bibinfo{volume}{81}},
  \bibinfo{pages}{041307} (\bibinfo{year}{2010}).

\bibitem[{\citenamefont{Lu et~al.}(2010)\citenamefont{Lu, Shan, Yao, Niu, and
  Shen}}]{Lu2010}
\bibinfo{author}{\bibfnamefont{H.-Z.} \bibnamefont{Lu}},
  \bibinfo{author}{\bibfnamefont{W.-Y.} \bibnamefont{Shan}},
  \bibinfo{author}{\bibfnamefont{W.}~\bibnamefont{Yao}},
  \bibinfo{author}{\bibfnamefont{Q.}~\bibnamefont{Niu}}, \bibnamefont{and}
  \bibinfo{author}{\bibfnamefont{S.-Q.} \bibnamefont{Shen}},
  \bibinfo{journal}{Phys. Rev. B} \textbf{\bibinfo{volume}{81}},
  \bibinfo{pages}{115407} (\bibinfo{year}{2010}).

\bibitem[{\citenamefont{Fu and Kane}(2008)}]{Fu2008}
\bibinfo{author}{\bibfnamefont{L.}~\bibnamefont{Fu}} \bibnamefont{and}
  \bibinfo{author}{\bibfnamefont{C.~L.} \bibnamefont{Kane}},
  \bibinfo{journal}{Phys. Rev. Lett.} \textbf{\bibinfo{volume}{100}},
  \bibinfo{pages}{096407} (\bibinfo{year}{2008}).

\bibitem[{\citenamefont{{Nilsson} et~al.}(2008)\citenamefont{{Nilsson},
  {Akhmerov}, and {Beenakker}}}]{Nilsson2008}
\bibinfo{author}{\bibfnamefont{J.}~\bibnamefont{{Nilsson}}},
  \bibinfo{author}{\bibfnamefont{A.~R.} \bibnamefont{{Akhmerov}}},
  \bibnamefont{and} \bibinfo{author}{\bibfnamefont{C.~W.~J.}
  \bibnamefont{{Beenakker}}}, \bibinfo{journal}{Phys. Rev. Lett.}
  \textbf{\bibinfo{volume}{101}}, \bibinfo{pages}{120403}
  (\bibinfo{year}{2008}).

\bibitem[{\citenamefont{{Akhmerov} et~al.}(2009)\citenamefont{{Akhmerov},
  {Nilsson}, and {Beenakker}}}]{Akhmerov2009}
\bibinfo{author}{\bibfnamefont{A.~R.} \bibnamefont{{Akhmerov}}},
  \bibinfo{author}{\bibfnamefont{J.}~\bibnamefont{{Nilsson}}},
  \bibnamefont{and} \bibinfo{author}{\bibfnamefont{C.~W.~J.}
  \bibnamefont{{Beenakker}}}, \bibinfo{journal}{Phys. Rev. Lett.}
  \textbf{\bibinfo{volume}{102}}, \bibinfo{pages}{216404}
  (\bibinfo{year}{2009}).

\bibitem[{\citenamefont{{Fu} and {Kane}}(2009)}]{Fu2009}
\bibinfo{author}{\bibfnamefont{L.}~\bibnamefont{{Fu}}} \bibnamefont{and}
  \bibinfo{author}{\bibfnamefont{C.~L.} \bibnamefont{{Kane}}},
  \bibinfo{journal}{Phys. Rev. Lett.} \textbf{\bibinfo{volume}{102}},
  \bibinfo{pages}{216403} (\bibinfo{year}{2009}).

\bibitem[{\citenamefont{Law et~al.}(2009)\citenamefont{Law, Lee, and
  Ng}}]{Law2009b}
\bibinfo{author}{\bibfnamefont{K.~T.} \bibnamefont{Law}},
  \bibinfo{author}{\bibfnamefont{P.~A.} \bibnamefont{Lee}}, \bibnamefont{and}
  \bibinfo{author}{\bibfnamefont{T.~K.} \bibnamefont{Ng}},
  \bibinfo{journal}{Phys. Rev. Lett.} \textbf{\bibinfo{volume}{103}},
  \bibinfo{pages}{237001} (\bibinfo{year}{2009}).

\bibitem[{\citenamefont{Tanaka et~al.}(2009)\citenamefont{Tanaka, Yokoyama, and
  Nagaosa}}]{TanakaYokoyama2009}
\bibinfo{author}{\bibfnamefont{Y.}~\bibnamefont{Tanaka}},
  \bibinfo{author}{\bibfnamefont{T.}~\bibnamefont{Yokoyama}}, \bibnamefont{and}
  \bibinfo{author}{\bibfnamefont{N.}~\bibnamefont{Nagaosa}},
  \bibinfo{journal}{Phys. Rev. Lett.} \textbf{\bibinfo{volume}{103}},
  \bibinfo{pages}{107002} (\bibinfo{year}{2009}).

\bibitem[{\citenamefont{{Zhou} et~al.}(2008)\citenamefont{{Zhou}, {Lu}, {Chu},
  {Shen}, and {Niu}}}]{Zhou2008}
\bibinfo{author}{\bibfnamefont{B.}~\bibnamefont{{Zhou}}},
  \bibinfo{author}{\bibfnamefont{H.}~\bibnamefont{{Lu}}},
  \bibinfo{author}{\bibfnamefont{R.}~\bibnamefont{{Chu}}},
  \bibinfo{author}{\bibfnamefont{S.}~\bibnamefont{{Shen}}}, \bibnamefont{and}
  \bibinfo{author}{\bibfnamefont{Q.}~\bibnamefont{{Niu}}},
  \bibinfo{journal}{Phys. Rev. Lett.} \textbf{\bibinfo{volume}{101}},
  \bibinfo{pages}{246807} (\bibinfo{year}{2008}).

\bibitem[{\citenamefont{Stanescu et~al.}()\citenamefont{Stanescu, Sau, Lutchyn,
  and Sarma}}]{Sarma2010}
\bibinfo{author}{\bibfnamefont{T.}~\bibnamefont{Stanescu}},
  \bibinfo{author}{\bibfnamefont{J.}~\bibnamefont{Sau}},
  \bibinfo{author}{\bibfnamefont{R.}~\bibnamefont{Lutchyn}}, \bibnamefont{and}
  \bibinfo{author}{\bibfnamefont{S.~D.} \bibnamefont{Sarma}},
  \bibinfo{note}{arXiv:1002.0842}.

\bibitem[{\citenamefont{Andreev}(1964)}]{Andreev1964}
\bibinfo{author}{\bibfnamefont{A.}~\bibnamefont{Andreev}},
  \bibinfo{journal}{Sov. Phys. JETP} \textbf{\bibinfo{volume}{19}},
  \bibinfo{pages}{1228} (\bibinfo{year}{1964}).

\bibitem[{\citenamefont{Blonder et~al.}(1982)\citenamefont{Blonder, Tinkham,
  and Klapwijk}}]{Blonder1982}
\bibinfo{author}{\bibfnamefont{G.~E.} \bibnamefont{Blonder}},
  \bibinfo{author}{\bibfnamefont{M.}~\bibnamefont{Tinkham}}, \bibnamefont{and}
  \bibinfo{author}{\bibfnamefont{T.}~\bibnamefont{Klapwijk}},
  \bibinfo{journal}{Phys. Rev. B} \textbf{\bibinfo{volume}{25}},
  \bibinfo{pages}{4515} (\bibinfo{year}{1982}).

\bibitem[{\citenamefont{Ou et~al.}(1987)\citenamefont{Ou, Hong, and
  Mandel}}]{Ou1987}
\bibinfo{author}{\bibfnamefont{Z.~Y.} \bibnamefont{Ou}},
  \bibinfo{author}{\bibfnamefont{C.~K.} \bibnamefont{Hong}}, \bibnamefont{and}
  \bibinfo{author}{\bibfnamefont{L.}~\bibnamefont{Mandel}},
  \bibinfo{journal}{Phys. Rev. A} \textbf{\bibinfo{volume}{36}},
  \bibinfo{pages}{192} (\bibinfo{year}{1987}).

\bibitem[{\citenamefont{Levitov et~al.}(1996)\citenamefont{Levitov, Lee, and
  Lesovik}}]{Levitov1996}
\bibinfo{author}{\bibfnamefont{L.}~\bibnamefont{Levitov}},
  \bibinfo{author}{\bibfnamefont{H.}~\bibnamefont{Lee}}, \bibnamefont{and}
  \bibinfo{author}{\bibfnamefont{G.}~\bibnamefont{Lesovik}},
  \bibinfo{journal}{J. Math. Phys.} \textbf{\bibinfo{volume}{37}},
  \bibinfo{pages}{4845} (\bibinfo{year}{1996}).

\end{thebibliography}

\end{document}